\documentclass[10pt,letterpaper]{article}
\usepackage[top=0.85in,left=2.75in,footskip=0.75in,marginparwidth=2in]{geometry}

\usepackage[utf8]{inputenc}
\usepackage{upgreek}
\usepackage{cite}

\usepackage{nameref,hyperref}

\usepackage{microtype}
\DisableLigatures[f]{encoding = *, family = * }

\raggedright
\usepackage{changepage}

\usepackage[aboveskip=1pt,labelfont=bf,labelsep=period,singlelinecheck=off]{caption}

\makeatletter
\renewcommand{\@biblabel}[1]{\quad#1.}
\makeatother

\usepackage{lastpage,fancyhdr,graphicx}
\usepackage{epstopdf}
\fancyheadoffset[L]{2.25in}
\fancyfootoffset[L]{2.25in}

\usepackage{color}

\definecolor{Gray}{gray}{.25}

\usepackage{graphicx}

\usepackage{sidecap}

\usepackage{wrapfig}
\usepackage[pscoord]{eso-pic}
\usepackage[fulladjust]{marginnote}
\reversemarginpar

\begin{document}
\vspace*{0.35in}

\begin{flushleft}
{\Large
\textbf\newline{Structural and electrical properties of fiber textured and epitaxial molybdenum thin films prepared by magnetron sputter epitaxy}
}
\newline
\\
Author Balasubramanian Sundarapandian\textsuperscript{1,*},
Author Mohit Raghuwanshi\textsuperscript{1},
Author Patrik Straňák\textsuperscript{1},
Author Yuan Yu\textsuperscript{1},
Author Haiyan Lyu\textsuperscript{2},
Author Mario Prescher\textsuperscript{2},
Author Lutz Kirste\textsuperscript{1}
Author Oliver Ambacher\textsuperscript{3}
\\
\bigskip
\bf{1} Fraunhofer Institute for Applied Solid State Physics, Tullastraße 72, 79108 Freiburg im Breisgau, Germany
\\
\bf{2} Institute of Physics (IA), RWTH Aachen University, Sommerfeldstraße 14, 52074 Aachen, Germany
\\
\bf{3} Institute for Sustainable Systems Engineering (INATECH), University of Freiburg, Emmy-Noether-Straße 2, University of Freiburg, 79110 Freiburg im Breisgau, Germany
\\
\bigskip
* balasubramanian.sundarapandian@iaf.fraunhofer.de

\end{flushleft}

\section*{Abstract}
Molybdenum (Mo) due to its optimal structural, physical, and acoustic properties find application as electrode material in aluminum scandium nitride (AlScN) and aluminum nitride (AlN)  based bulk acoustic wave (BAW) resonators. Epitaxial Mo thin films exhibiting low resistivity can improve the performance of the BAW resonator by enhancing both the electromechanical coupling coefficient ($k^2_\mathrm{eff}$) and quality factor \textit{Q}. In this study, we systematically vary the growth temperature of Mo grown on fiber-textured and epitaxial wurtzite-aluminum nitride (AlN) to study the changes in structural and electrical properties of the Mo films. Results show that Mo grown at 700°C on epitaxial AlN exhibit low surface roughness ($R_\mathrm{q}$ = 0.8$\,$nm), large average grain diameter ($d_\mathrm{grain}$ = 330$\,$nm), low resistivity ($\rho$ = 6.6$\,\pm\,$0.06$\,\upmu\Omega$cm), and high crystal quality (XRD Mo 110 $\upomega$-FWHM = 0.63°). XRD pole figure and $\phi$-scan \mbox{analysis} reveal that irrespective of the growth temperature, Mo is fiber textured on \mbox{fiber-textured} AlN, and has three rotational domains on epitaxial AlN. The study shows that the resistivity of Mo reduces with increasing growth temperature, which we relate to increasing average grain diameter. Additionally, we show that fiber-textured Mo has more high angle grain boundaries resulting in consistently higher resistivity than its epitaxial equivalent.


The advent of 5G telecommunication standard demands wide bandwidth filters for operation at frequencies ranging between 3.3$\,$GHz and 5.9$\,$GHz with minimal acoustic loss.\cite{aigner2018baw} Bulk acoustic wave (BAW) resonators offer high electro-mechanical coupling coefficient ($k^2_\mathrm{eff}$),  high quality factor (\textit{Q}), and low temperature coefficient of frequency (TCF), making them desirable for the requirements imposed by the telecommunication industry.\cite{nishihara2002high,mahon20175g,aigner2008saw,malocha2010saw} BAW resonators consist of a piezoelectric material sandwiched between electrodes, material properties of these layers determine the figure of merit (FOM) [$k^2_\mathrm{eff}\,\times\,$\textit{Q}] of these resonators.
Aluminum nitride (AlN) thin films due to their optimal piezoelectric, \cite{naik2000measurements,tonisch2006piezoelectric} acoustic, \cite{dubois1999properties} and physical properties \cite{slack1987intrinsic} were conventionally used as piezoelectric material. After the discovery that aluminum scandium nitride\cite{akiyama2009enhancement} (AlScN) prepared by alloying AlN and scandium nitride (ScN) can provide relatively high intrinsic electro-mechanical coupling coefficient $k^2_\mathrm{t}$ and  piezo-electric coefficient $d_{33}$,\cite{wingqvist2010increased} it is replacing AlN based BAW resonators. BAW resonators employing epitaxial AlN as the piezoelectric material, showed improved power handling capabilities,\cite{vetury2018high} and $k^2_\mathrm{eff}$.\cite{shealy2016single}  The same is expected for AlScN, therefore a bottom electrode material exhibiting epitaxy is desired to promote epitaxial growth of AlScN. Materials such as platinum (Pt),\cite{dubois1999baw} molybdenum (Mo),\cite{cherng2011pulsed} iridium (Ir),\cite{clement2009aln} and tungsten (W) \cite{bradley200712e} are commonly used as electrodes for AlN based BAW resonators. They have shown to promote AlN growth in the 0001 direction, hence the same is expected for wurtzite-AlScN. Additionally, properties such as acoustic impedance $z_\mathrm{ac}$, electrical resistivity $\rho$, and low mass density $\rho _\mathrm{mass}$ determine the performance of the BAW resonators. Large differences in $z_\mathrm{ac}$ of two media increases the reflection of waves at the interfaces and  therefore may increase the energy density leading to an increased $k^2_\mathrm{eff}$ in acoustic resonators. Therefore, a large disparity in acoustic impedance of the piezoelectric material and electrode material is desired. Low $\rho$, and low $\rho _\mathrm{mass}$ is particularly crucial to improve \textit{Q} and reduce mass loading, \cite{thalhammer2006ohmic,knapp2018graphene} respectively. Lee \textit{et al.} have studied the influence of electrode materials such as molybdenum (Mo), aluminum (Al), and tungsten (W) on the resonator properties and found that FOM for Mo based solidly mounted resonator (SMR) to be higher than Al and W based resonators.\cite{lee2002influence} Ueda \textit{et al}. showed that Mo, Ru, and W based film bulk acoustic resonators (FBARs) achieved low losses by comparing FBARs with electrode materials such as Al, titanium (Ti), copper (Cu), gold (Au), chromium (Cr), Mo, ruthenium (Ru), and W.\cite{ueda2007film} Since, Mo was found to have optimal structural, physical, and acoustic properties, it has emerged as the choice of electrode material for BAW resonators. To promote oriented growth of Mo on Si substrates, other groups have used AlN as a seed layer,\cite{kamohara2005growth, kamohara2008influence, cherng2011pulsed} therefore a similar approach was taken in this work. Additionally, the thermal expansion coefficient of Mo is
similar to AlN, \cite{yim1974thermal, hidnert1924thermal} making AlN an ideal seed layer for Mo growth. In this work, we systematically vary the growth temperature and demonstrate the change in structural and electrical properties of Mo thin films, sputtered on fiber-textured and epitaxial AlN thin films prepared by magnetron sputter epitaxy.

AlN and Mo films were prepared using an Evatec Clusterline 200II planar magnetron sputtering module with a leak rate of $\approx\,$7.5$\,\times\,$10$^{-7}$ mbar l/s. Both the films were deposited at a working pressure of $\approx\,$1$\,\times\,$10$^{-3}$ mbar. Al, and Mo targets (304$\,$mm diameter) were used to deposit AlN, and Mo films onto 200$\,$mm Si(111) substrates. The native oxide layer that grows unintentionally on the silicon substrate was cleaned using in-situ inductively coupled plasma (ICP) etching. All the substrates were etched with an MF power of 500$\,$W, RF power of 50$\,$W, at $\approx\,$5$\,\times\,$10$^{-6}$ mbar Ar pressure for 85$\,$s. More information about thickness of native oxide and its evolution after ICP etching can be found in our previous work.\cite{weippert2024formation} AlN was sputtered at 500$\,$°C, and 700$\,$°C to obtain fiber-textured and epitaxial films, respectively.\cite{sundarapandian2023influence} Process parameters such as power, gas flow, and target to substrate distance (TSD) were kept constant at 5500$\,$W, 40$\,$sccm, and 10$\,$mm respectively. Mo thin films were sputtered on AlN thin films at growth temperatures between 300$\,$°C and 700$\,$°C in steps of 100$\,$°C, while keeping the power, gas flow, and TSD constant at 1000$\,$W, 25$\,$sccm, and 60$\,$mm, respectively. Surface morphology, surface roughness ($R_\mathrm{q}$), and average grain diameter ($d_\mathrm{grain}$) of the films were characterized using Bruker ICON AFM. An IONTOF M6 Plus time-of-flight secondary ion mass spectrometer (TOF-SIMS) was used to determine the thickness of the films, and interface diffusion. A PANalytical X’Pert Pro MRD diffractometer with a Ge-220 hybrid monochromator providing Cu-K$\alpha _1$ was used to perform X-ray diffraction (XRD) and X-ray reflectivity (XRR) analysis. Electrical resistivity ($\rho$) of the films were determined at the center of the wafer using four point probe method. To account for the error, each sample was measured five times at approximately the same position. EBSD measurements were carried out using an EDAX-Hikari camera mounted on the FEI Helios NanoLab 650. EBSD analyses were performed under SEM conditions at 20 kV and 1.6 nA. Data acquisition utilized a 2$\,\times\,$2 bin-ning resolution and the subsequent data processing was completed using OIM analysis software, which identified properties of GBs based on the detected Kikuchi patterns.

\begin{figure}[!h]
	\centering
	\includegraphics[width=1\linewidth]{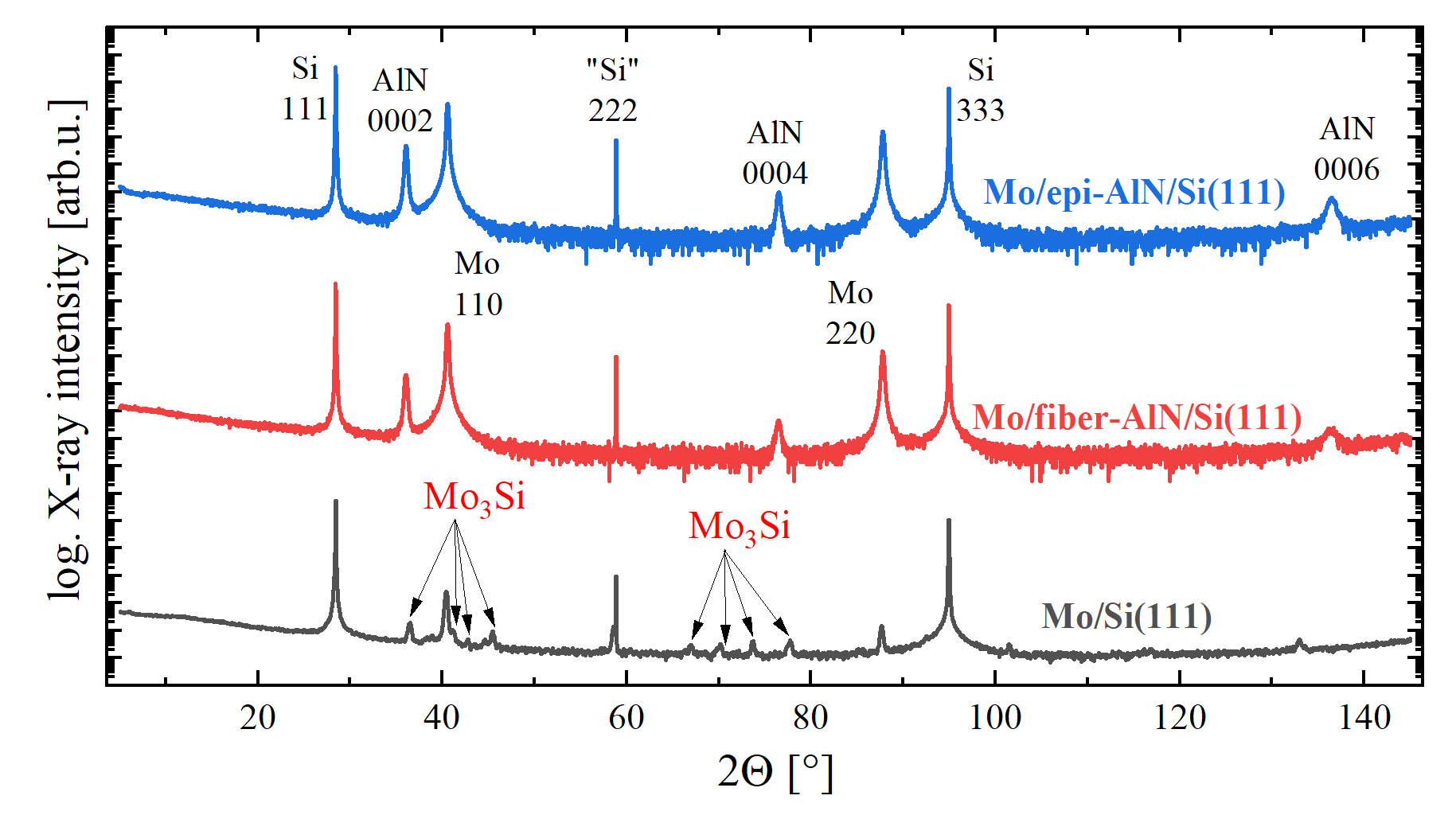}
\caption{XRD-$2\theta / \theta$ scans of molybdenum sputtered at 700°C on Si(111), fiber-textured AlN, and epitaxial AlN.}
\label{fig:th_2th_epi_fiber}
\end{figure}

XRD $2\theta / \theta$ scans performed on Mo deposited at 700°C on Si(111), fiber-textured, and epitaxial AlN are shown in figure \ref{fig:th_2th_epi_fiber}. The scans demonstrate absence of peaks corresponding to molybdenum silicides (Mo$_3$Si) when body-centered cubic (bcc) Mo was sputtered on fiber-textured and epitaxial AlN. This result emphasizes the importance of the use of AlN as a seed layer for Mo growth on silicon substrates. Peaks from Mo$_3$Si were also absent when Mo was sputtered at growth temperatures below 700°C (see figure S1 in supplementary material). Additionally, ToF-SIMS measurement showed that there is slight diffusion of Mo, and Si into the AlN layer (see figure S5 in supplementary), indicating that a certain thickness of AlN layer is required to prevent formation of Mo$_3$Si. The thickness of the films (see table S1), as determined using ToF-SIMS measurements remained constant for all growth temperatures at $\approx\,$120$\,$nm, and hence the growth rate at $\approx\,$1.04$\,$nm/s. Crystal orientations that can be observed for Mo deposited on AlN depends on the polarity of AlN. When bcc-Mo is deposited on N-Polar AlN, Mo is expected to grow with both (110) and (100) planes parallel to the AlN (0001) basal plane, while on metal polar AlN only Mo (110) plane is expected.\cite{gao2010cube} For Mo deposited on fiber-textured and epitaxial AlN, reflections from AlN 000\textit{l} (\textit{l} = 2,4,6), Mo 110 and Mo 220 were observed. This shows that Mo grows with its (110) plane parallel to the AlN (0001) basal plane. In our previous work, we have shown that our AlN is mixed polar\cite{sundarapandian2023influence,sundarapandian2024comparison} which could be the reason for the absence of Mo 200 reflection. Moreover, $2\theta / \theta$ analysis of a similar stack deposited on sapphire showed reflection from (200) plane.\cite{wolff2021atomic} Since (110) planes of Mo are densely packed, they are expected to have minimum surface, and interface energies, promoting (110) plane parallel to the substrate.\cite{thompson1990grain} Given the preference, the (100) planes that are expected to grow on N-polar domains maybe are suppressed, leading to the absence of (100) planes parallel to the substrate.

\begin{figure}[!h]
	\centering
	\includegraphics[width=1\linewidth]{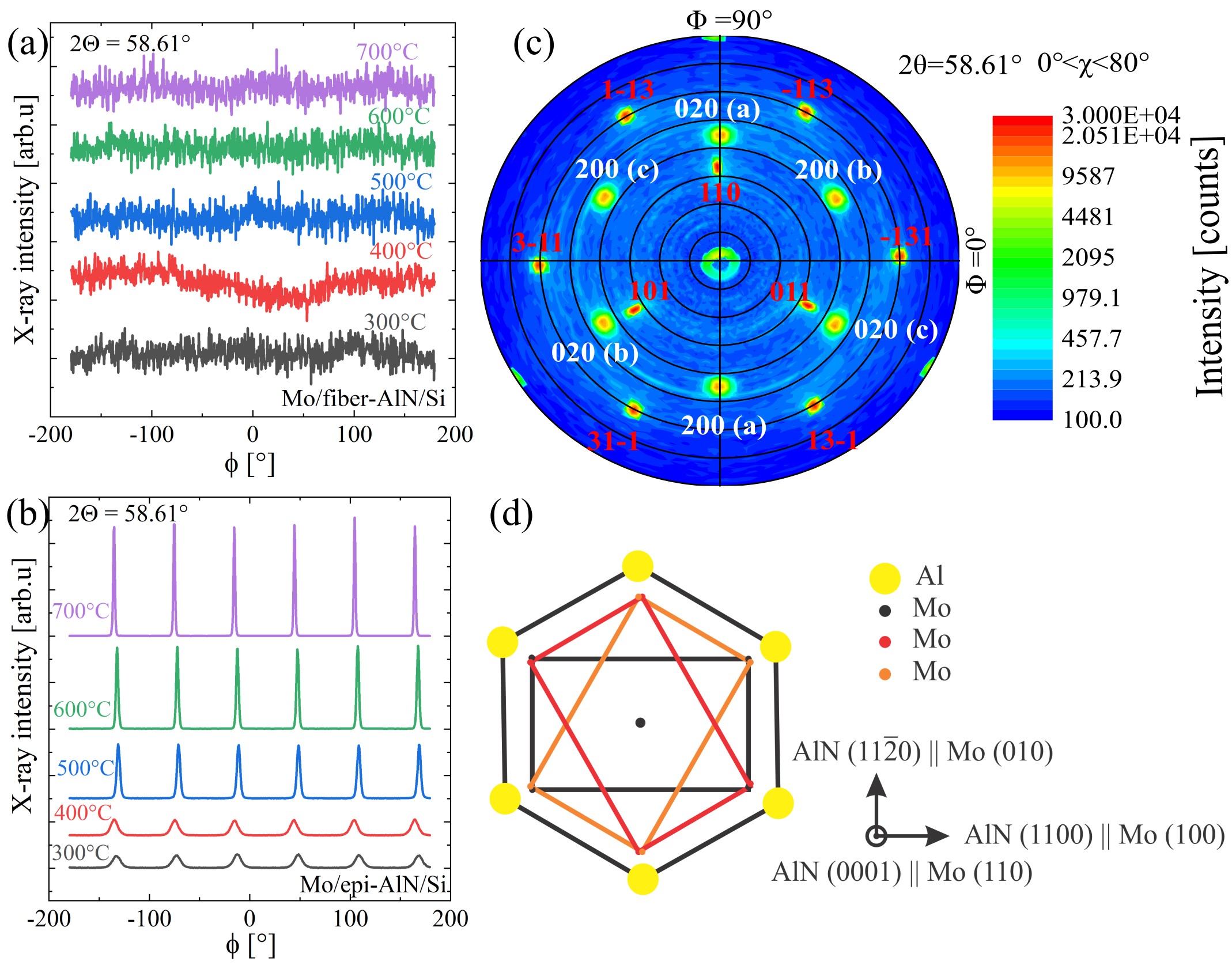}
\caption{$\phi$-scan of Mo 200 reflection recorded from molybdenum deposited on (a) fiber-textured AlN, (b) Epitaxial AlN at various growth temperatures. (c) pole figure recorded at 2$\theta\,$=58.61° for Mo deposited at 700°C on epitaxial AlN (reflections marked in red correspond to Si, while the ones in white correspond to Mo ), and (d) Schematic demonstrating the three domain structure of Mo on AlN (0001) basal plane.}
\label{fig:mo_phi_pol}
\end{figure}

$\phi$-scan, and pole figure (PF) shown in figure \ref{fig:mo_phi_pol} were recorded for Mo 200  reflection at $2\theta\,\approx\,$58.61°. Irrespective of the growth temperature, Mo deposited on fiber-textured AlN showed no peaks in $\phi$-scan indicating that Mo is fiber-textured \cite{kamohara2008improvement} (see figure \ref{fig:mo_phi_pol} (a)). On epitaxial AlN, at all growth temperatures $\phi$-scan recorded from Mo 200 reflection showed six peaks separated by $\approx\,$60°, indicating that Mo has an orientational relationship (OR) with AlN  (see figure \ref{fig:mo_phi_pol}(b)). This result also shows that the texture of Mo is independent of the growth temperature, and is dictated predominantly by the crystallographic texture of AlN seed layer. Distribution of grain misorientation determined using EBSD  (see figure S4 in supplementary material) showed that the grains prefer to grow with a misorientation angle of $\approx\,$60° irrespective of the crystallographic texture, as this orientation has been observed to have the least grain boundary energy in bcc metals.\cite{chirayutthanasak2024universal} Even though there is a preferred orientation for the grains, it is possible that the interface energy anisotropy for grain rotation in case of epitaxial AlN is high, constraining Mo grains to grow in a specific orientation.\cite{thompson1990grain} PF recorded between $\chi$ angles of 0° and 80° is shown in figure \ref{fig:mo_phi_pol} (c), reflections marked in red correspond to silicon while the ones marked in white correspond to Mo. At $\chi \approx \,$45°, six reflections can be observed, matching the number of reflections seen in the $\phi$-scans. Stereographic projections  simulated for Mo oriented along the [110] direction showed only two reflections corresponding to (100) and (010) planes at $\chi\,$=$\,$45°. The presence of additional four reflection in the PF indicates the presence of additional two domains of Mo which was also observed in EBSD maps (see figure s3 in supplementary material). The reflections from these three domains are marked with letters a, b, and c in figure \ref{fig:mo_phi_pol} (c). Similar conclusions were made for bcc$\Vert$hcp systems.\cite{meijers2005epitaxial, gao2009thermal,wolff2021atomic} The three domains of Mo(110) on AlN(0001) basal plane is schematically shown in figure \ref{fig:mo_phi_pol} (d), the three domains are marked in black, red, and orange. The OR of the three domains were determined to be AlN[11$\overline{2}$0]$\Vert$Mo[010], AlN[$\overline{2}$110]$\Vert$Mo[010] and AlN[1$\overline{2}$10]$\Vert$Mo[010]. Since Mo on fiber-textured AlN is fiber-textured, and has an OR with epitaxial AlN, they will be called ``Fiber-Mo'' and ``Epi-Mo'', respectively in the rest of the text. At the measured 2$\theta$ angle, reflections from Si\{113\} and Si\{110\} were observed at $\chi\,\approx\,$60° and $\chi\,\approx\,$35° respectively. These reflections are marked red in figure \ref{fig:mo_phi_pol} (c). Table \ref{tab:soa} compares present study's findings with previously reported XRD rocking curve full width at half maximum ($\omega$-FWHM) of Mo 110 reflection sputtered on Si with an AlN seed layer. The table clearly demonstrates that the crystalline properties of Mo reported in this work is superior to the previously reported values. Based on the information available in the papers, and given that we obtain Mo with high crystalline quality on both fiber textured and epitaxial AlN at 700°C, we believe that growth temperature is a key factor in obtaining Mo with high crystalline quality on AlN.

\begin{table}[!h]
\caption{Table comparing reported values of $\omega$-FWHM of Mo 110 with this work}
\resizebox{\columnwidth}{!}{
\begin{tabular}{cccc}
\hline \hline
\textbf{}                                                     & \textbf{\begin{tabular}[c]{@{}c@{}}Thickness\\ AlN seed/Mo\\ {[}nm{]}\end{tabular}} & \textbf{\begin{tabular}[c]{@{}c@{}}$\omega$-FWHM Mo 110\\ {[}°{]}\end{tabular}} & \textbf{Texture}                                                   \\ \hline
J.S. Cherng et al.\cite{cherng2011pulsed}                                            & 25/1500                                                                             & 1.9                                                                          & \begin{tabular}[c]{@{}c@{}}Information\\ not availabe\end{tabular} \\
R.E.Sah et al.\cite{sah2013crystallographic}                                                                                           & 600/170                                                                             & 6                                                                            & \begin{tabular}[c]{@{}c@{}}Information\\ not availabe\end{tabular} \\
Toshihiro et al.\cite{kamohara2008improvement}                                                                                         & 100/200                                                                             & 4                                                                            & \begin{tabular}[c]{@{}c@{}}Fiber\\ textured\end{tabular}           \\
\begin{tabular}[c]{@{}c@{}}This work (700°C Fiber-Mo)\end{tabular} & 85/116                                                                             & 0.78                                                                         & \begin{tabular}[c]{@{}c@{}}Fiber\\ textured\end{tabular}           \\
\begin{tabular}[c]{@{}c@{}}This work (700°C Epi-Mo) \end{tabular} & 97/116                                                                             & 0.63                                                                         & \begin{tabular}[c]{@{}c@{}}Three rotational domains\\ exhibiting OR with AlN \\
\end{tabular} \\ \hline \hline
\end{tabular}
}
\label{tab:soa}
\end{table}

\begin{figure}[!h]
	\centering
	\includegraphics[width=1\linewidth]{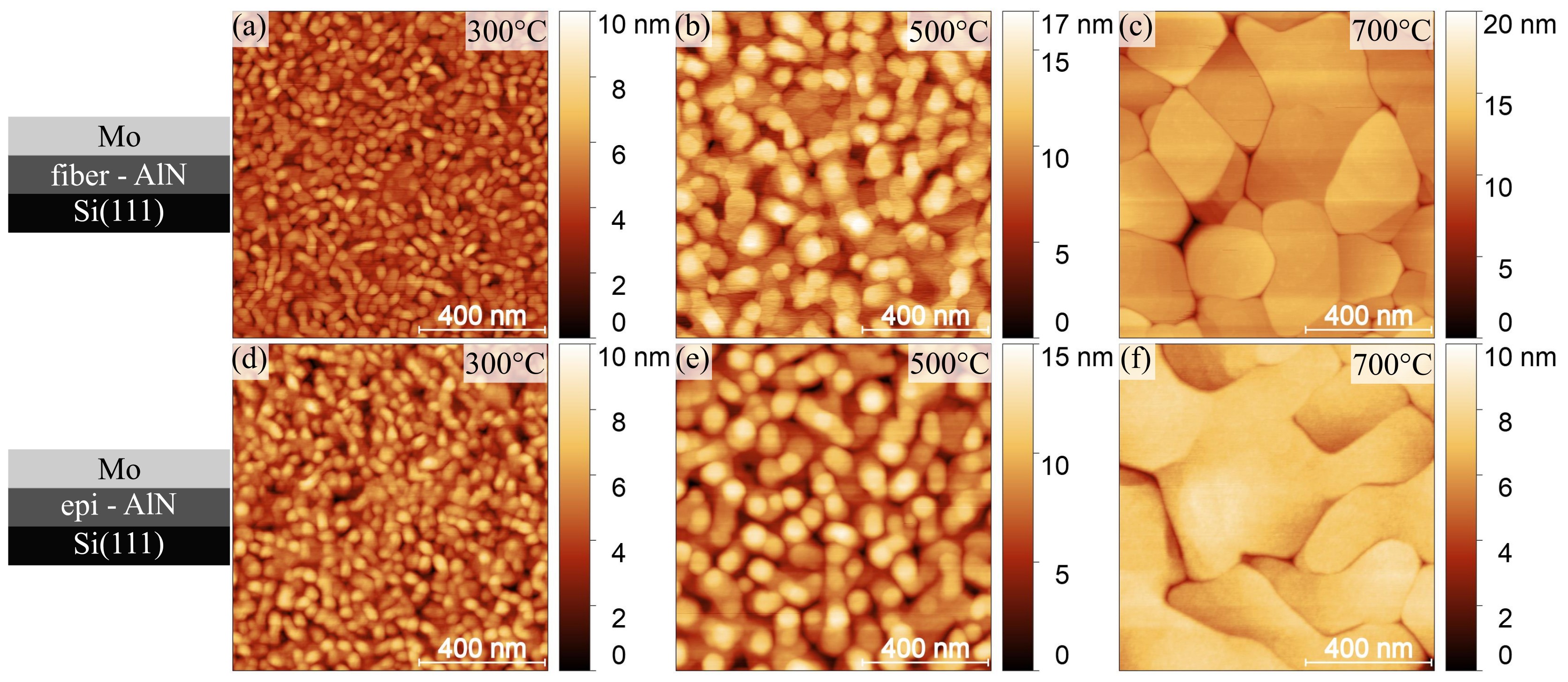}
\caption{Moybdenum grown on (a)-(c) fiber-textured and (d)-(f) on epitaxial AlN at growth temperatures of 300°C, 500°C and 700°C.}
\label{fig:afm_mo}
\end{figure}

\begin{figure}
    \centering
    \includegraphics[width=1\linewidth]{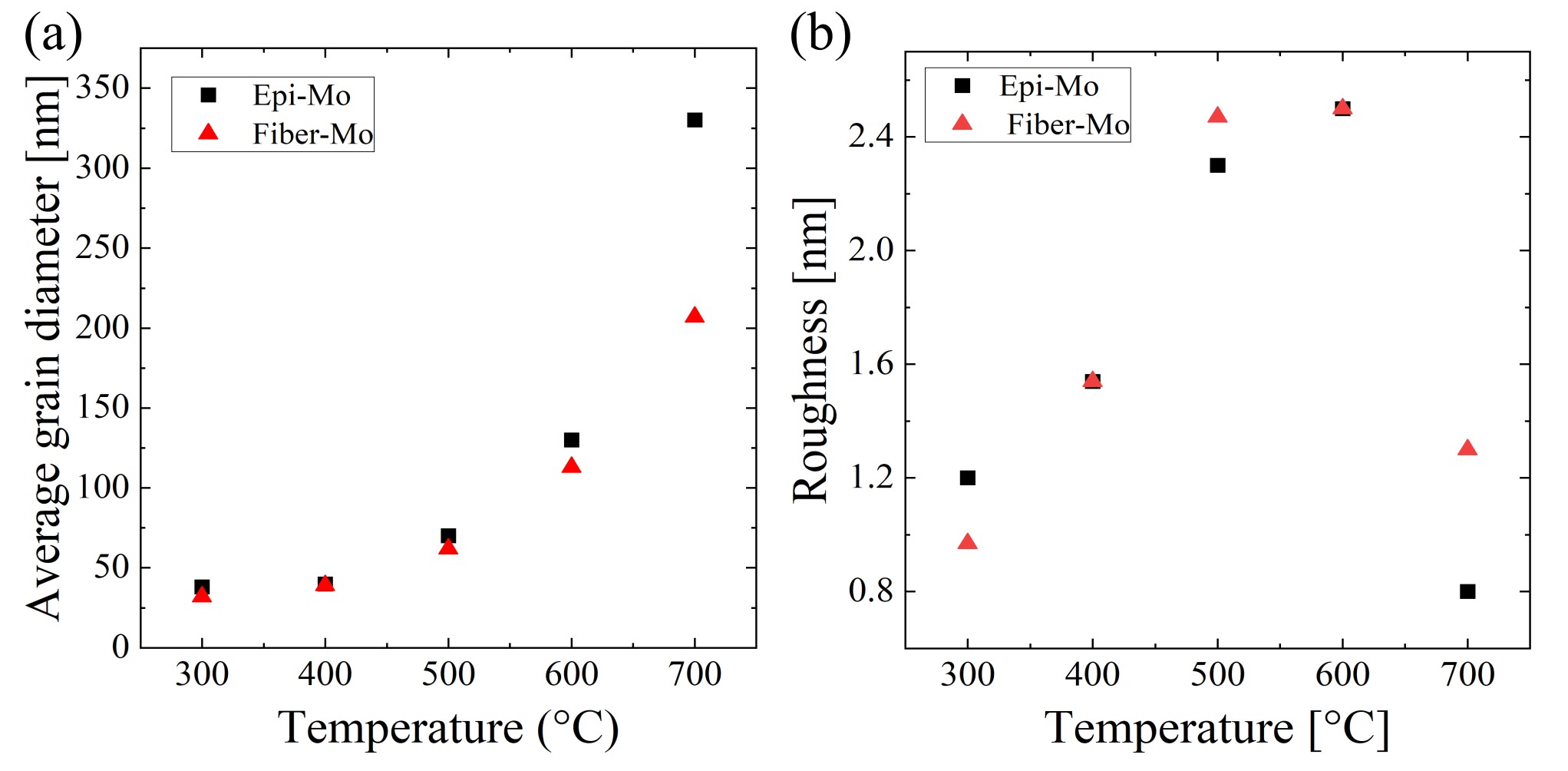}
    \caption{Change in (a) average grain diameter ($d_\mathrm{grain}$), and (b) RMS roughness ($R_\mathrm{q}$) of Epi and fiber Mo.}
    \label{fig:grain_size_roughness}
\end{figure}

The surface morphology of Mo deposited at growth temperatures of 300°C, 500°C, and 700°C on fiber-textured and epitaxial AlN is shown in figure \ref{fig:afm_mo}. Mo on fiber-textured AlN is shown in panels (a) - (c), while panels (d) - (f) show Mo grown on epitaxial AlN. As a consequence of the increase in adatom diffusion with temperature, an increase in grain diameter was observed in both fiber and epi-Mo.
The average grain diameter ($d_\mathrm{grain}$) determined from the AFM images is shown in figure \ref{fig:grain_size_roughness} (a), in both the cases $d_\mathrm{grain}$ increased with growth temperature. The surface diffusion coefficient \textit{D} can be expressed as $D_0e^{-E/kT}$,\cite{ohring2001materials} the exponential increase in \textit{D} with growth temperature could be the reason for the increase in $d_\mathrm{grain}$. For Epi-Mo, the $d_\mathrm{grain}$ increased from $\approx \,$38$\,$nm at 300°C to $\approx \,$330$\,$nm at 700°C, while for fiber-textured Mo $d_\mathrm{grain}$ increased from $\approx \,$32$\,$nm to $\approx \,$207$\,$nm. Above 500°C, the rate of increase in $d_\mathrm{grain}$ is less pronounced for fiber-textured Mo, indicating that the texture of Mo has an influence on the surface diffusion of ad-atoms. Since Mo is fiber-textured, the random orientation of Mo grains could act as an additional barrier for Mo adatoms to diffuse between grains, thereby limiting the $d_\mathrm{grain}$. 

\begin{figure}[!h]
	\centering
	\includegraphics[width=0.8\linewidth]{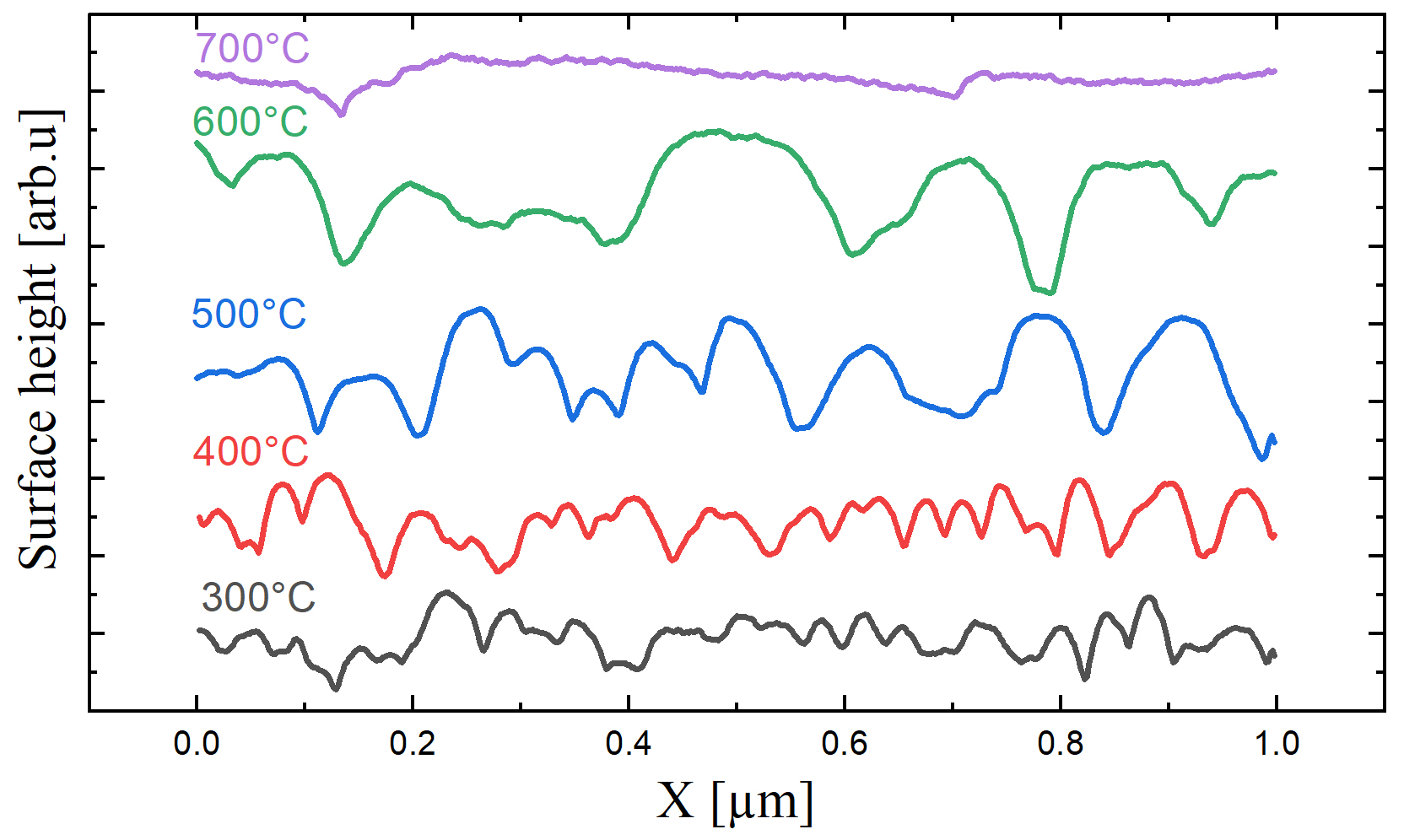}
\caption{Surface profile of epitaxial Mo deposited at various growth temperatures.}
\label{fig:surface_profile}
\end{figure}

The change in surface roughness ($R_\mathrm{q}$) as a function of growth temperature determined as root mean square (RMS) of the surface height for epitaxial and fiber-textured Mo is depicted in figure \ref{fig:grain_size_roughness}(b). Interestingly, $R_\mathrm{q}$ reached a maximum at 600°C, for both the cases. XRR recorded for these samples showed that the rate of decay of reflections which is roughness dependent,\cite{moram2009x} followed a similar trend (see figure S2 in supplementary file). Additionally, the mass density of the films extracted by fitting the reflections remained constant at 10.2$\,$g/cm$^3$, indicating that the density of Mo is independent of growth temperature, and crystallographic texture. For fiber-textured and epitaxial Mo, $R_\mathrm{q}$ dropped to 1.3$\,$nm and 0.8$\,$nm, respectively at 700°C. Surface profiles extracted from AFM images of epitaxial Mo deposited at different growth temperatures is shown in figure \ref{fig:surface_profile}. With increasing growth temperature, a change in shape of the grains can be observed. At growth temperatures of 300°C, and 400°C the surface of the grains predominantly have an inverted V-shape. The inverted V-shape is a result of anisotropy in growth rate of different crystallographic planes, the facets of the inverted V corresponds to crystallographic planes with low growth rates. As growth temperature increases, surface diffusion of ad-atoms between grains happen resulting in grain growth. At 600°C, the grains start to flatten, and at 700°C almost no inverted V shaped grains can be observed. This could be due to recrystallization resulting in the observed drop in $R_\mathrm{q}$. Since 600°C, corresponds to a $T_s$/$T_m$ ratio of 0.3, recrystallization can be expected at this temperature.\cite{shewmon1969transformations} Here, $T_m$ is the melting point of Mo, and it was assumed to be 2895$\,$K \cite{worthing1925temperature}, and $T_s$ is the substrate temperature. 

\begin{figure}[!h]
	\centering
	\includegraphics[width=1\linewidth]{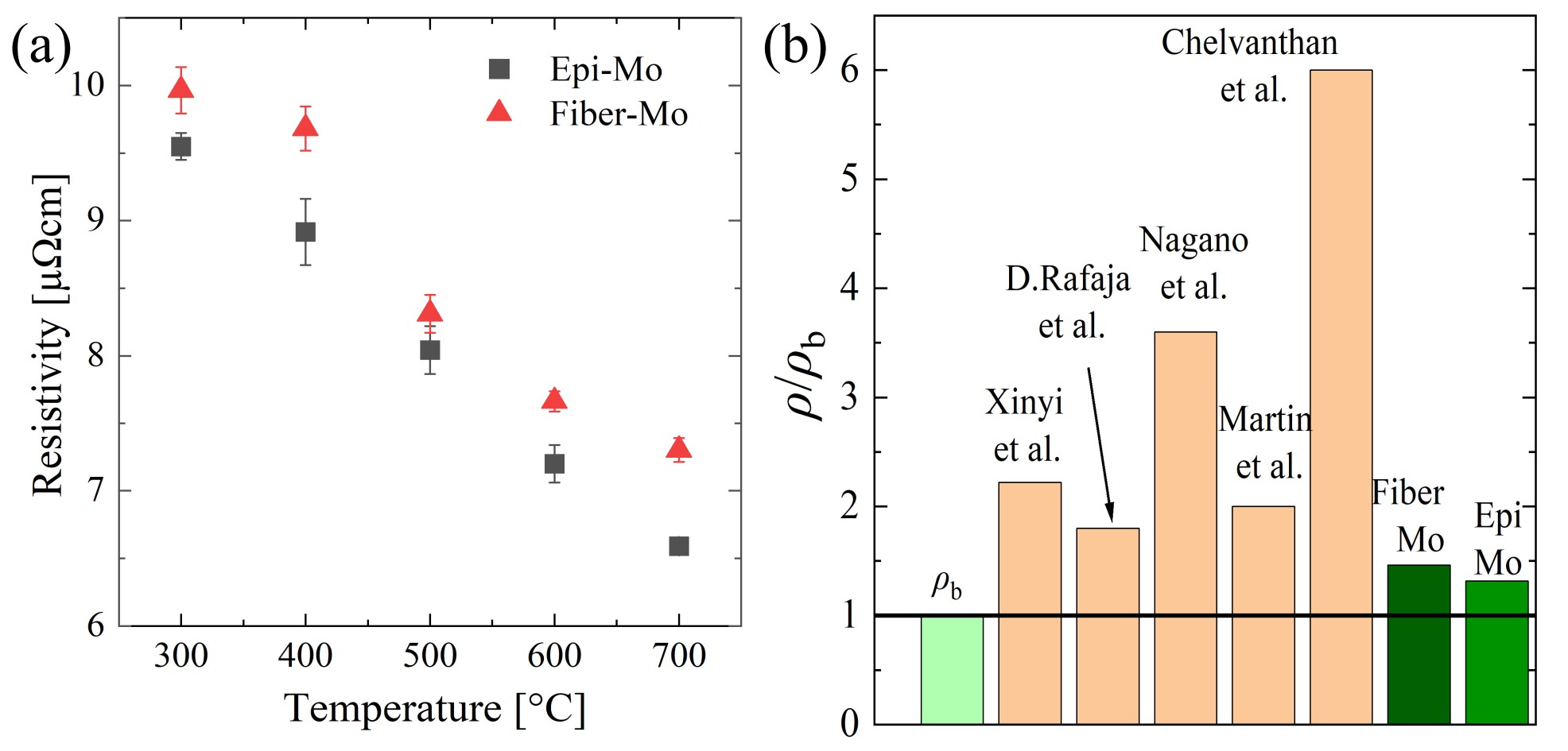}
\caption{(a) Resistivity ($\rho$) of epitaxial and fiber-textured Mo determined by four point probe method, and (b) Comparison of Mo resistivity from this work with other researchers and its bulk resistivity $\rho _\mathrm{b}$. }
\label{fig:resistivity}
\end{figure}

Electrical resistivity ($\rho$) of epi, and fiber Mo determined using four point probe method is shown in figure \ref{fig:resistivity} (a). $\rho$ of both epi, and fiber Mo decreased with increasing growth temperature. $\rho$ of thin films depends on the thickness of the films, grain boundary scattering, scattering due to macroscopic roughness, and impurities in thin films. $\rho$ is expected to decrease with increasing thickness, due to scattering of electrons at the film surface. Since the thickness of the Mo films investigated were approximately the same, it is not contributing to the observed decrease in $\rho$. Increase in roughness of the films is expected to increase the $\rho$.\cite{namba1970resistivity} Though roughness of both fiber-textured and epitaxial Mo increase up to a growth temperature of 600°C,  a decrease in $\rho$ can be observed. This shows that the contribution to $\rho$ by roughness, even if present is not the dominant effect. Oxygen concentration in Mo films can also affect its $\rho$, \cite{bardin1988effects} ToF-SIMS measurement revealed that the Mo thin films have approximately the same level of oxygen irrespective of its crystallographic texture, and growth temperature (see Figure S6). Therefore, oxygen concentration in the films have a negligible effect on the observed change in $\rho$. Grain boundaries can act as planes that partially reflect electrons,\cite{mayadas1970electrical} thereby increasing $\rho$ of the films. With increasing $d_\mathrm{grain}$, grain boundary density reduces and hence its $\rho$. In figure \ref{fig:grain_size_roughness} (a) it was shown that $d_\mathrm{grain}$ increases with temperature, which could be the reason for the decreasing $\rho$ with temperature. Figure \ref{fig:resistivity} (b), compares the resistivity of fiber Mo and epi Mo obtained in this work with work of other researchers. Here, the resistivity is normalized to bulk resistivity of Mo $\rho _\mathrm{b}$. $\rho _b$ was assumed to be 5.5$\,\upmu\Omega$cm.\cite{white1959electrical} From the figure, it can be seen that the resistivity of Mo obtained in this work is close to $\rho _\mathrm{b}$ and better than other works. \cite{dai2014molybdenum, rafaja2013effect, martin2006process, chelvanathan2015annealing, nagano1980electrical}

\begin{figure}[!h]
	\centering
	\includegraphics[width=0.75\linewidth]{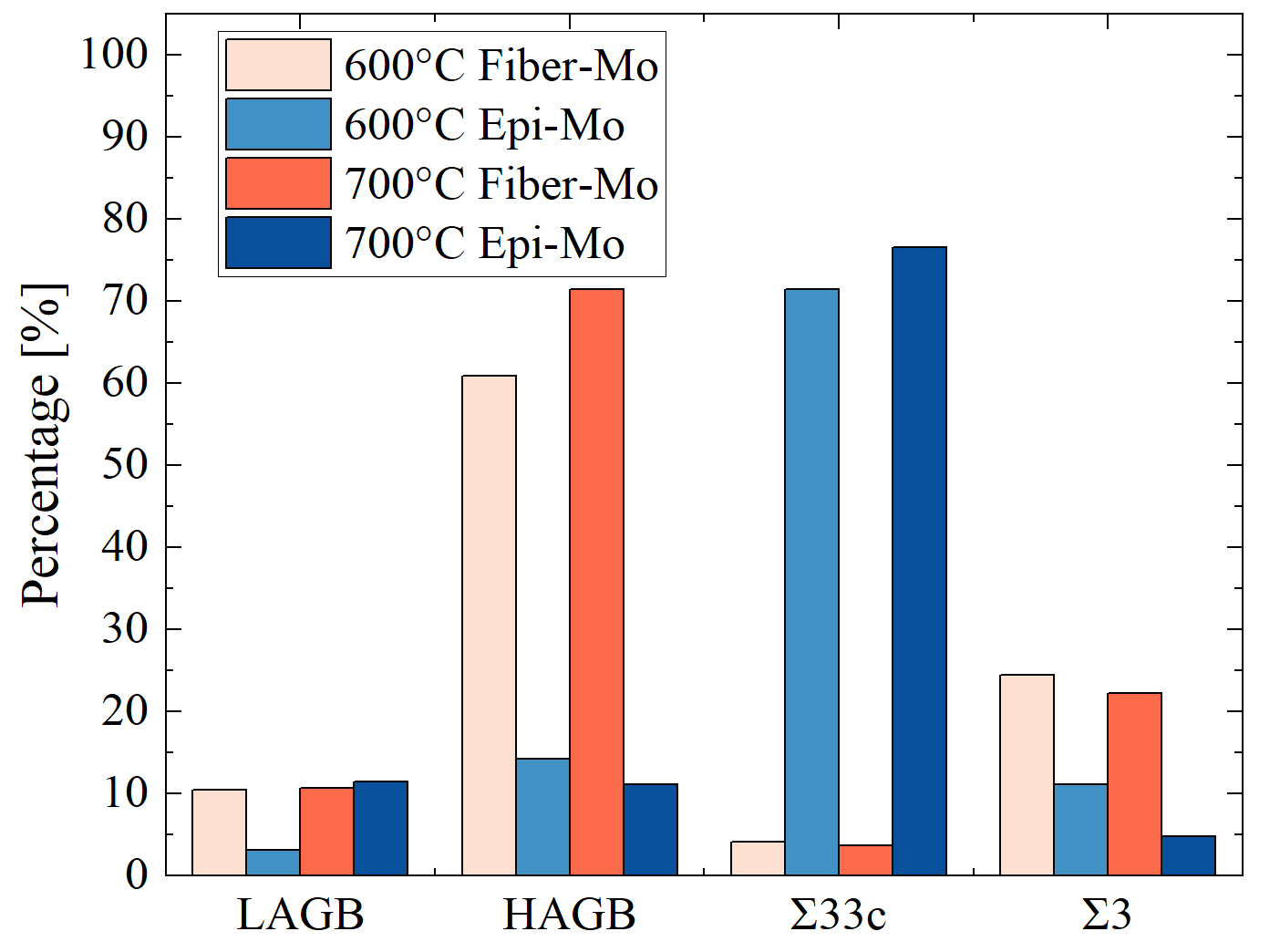}
\caption{Percentage of low angle, high angle, $\Sigma$3, and $\Sigma$33c grain boundaries present in molybdenum grown on fiber-textured and epitaxial AlN at growth temperatures of 600°C, and 700°C.}
\label{fig:gb}
\end{figure}

At growth temperatures below 600°C, where the $d_\mathrm{grain}$ of both epitaxial and fiber-textured Mo is approximately the same, epitaxial Mo showed consistently lower $\rho$ than fiber-textured Mo (see figure \ref{fig:resistivity} (a)). This indicates that there is an additional effect contributing to the $\rho$ of the films. In addition to grain diameter, the type of grain boundary \textit{i.e} whether the grain boundary is a low-angle grain boundary(LAGB), high-angle grain boundary (HAGB), or a sigma ($\Sigma$) boundary influences the scattering of electrons at the grain boundary.\cite{bishara2020approaches,bishara2021understanding,cesar2014calculated,lee2018first,zhou2022first} Therefore, the grain boundaries present in Mo films grown at 600°C and 700°C on fiber-textured and epitaxial AlN were characterized using EBSD. The analysis showed that all the films contained LAGBs, HAGBs, $\Sigma$33c, and $\Sigma$3 grain boundaries, the percentage distribution of these boundaries in the films is shown in figure \ref{fig:gb}. The results clearly show that Mo grown on epitaxial AlN has more $\Sigma$ boundaries than fiber-textured Mo, while fiber-textured Mo has significantly more HAGB than epitaxial Mo. In Mo, HAGB are more resistive than LAGB, \cite{karolik1994calculation} and since $\Sigma$ boundaries have high symmetry one could expect low scattering of electrons at these boundaries than HAGB. Therefore, it can be concluded that the grain boundaries in fiber-textured Mo scatters more electrons than epitaxial Mo, resulting in higher $\rho$ at growth temperatures below 600°C. Furthermore, an increase in grain boundary density increases the impurities concentrated at the grain boundaries. This in turn increases the potential barrier for mobile electrons to move from one grain to another, localizing electrons in their own grains,\cite{vancea1982reduced, nguyen2018electron} increasing $\rho$ of the films. A systematic study is required to characterize the impurities at the grain boundaries to further comment on this. 

In summary, our results show that the growth temperature of Mo has significant effects on the, crystal quality, average grain diameter, roughness, and electrical resistivity of the films. Through XRD $\phi$-scan, and pole figure analysis we show that the crystallographic texture of Mo sputtered on AlN thin films, is dictated by the texture of the AlN film, independent of growth temperature. Irrespective of the crystallographic texture of Mo, the average grain diameter increases with growth temperature due to increased mobility of ad-atoms. The roughness of the films reach a maximum at a growth temperature of 600°C, and drops significantly at 700°C, due to re-crystallization of Mo. The resistivity of the films decrease with increasing growth temperature, as a consequence of increasing grain diameter, reaching a minimum of 6.6$\,\pm\,$0.06$\,\upmu\Omega$cm at 700°C for Mo grown on epitaxial AlN. Fiber-textured Mo has more high angle grain boundaries than epitaxial Mo, leading to increased resistivity of these films. Our results show that Mo grown at 700°C on epitaxial AlN has favourable properties for use as electrode material in AlScN and AlN based BAW resonators.

\section*{Supplementary material}
See supplementary material for XRD-$2\theta / \theta$ scans of Mo sputtered at various temperatures on fiber textured and epitaxial AlN (Figure S1), XRR scans recorded for fiber textured and epitaxial Mo(Figure S2), exemplary EBSD map of fiber-Mo and epi-Mo (Figure S3), distribution of grain misorientation in fiber-Mo, and epi-Mo (Figure S4), ToF-SIMS measurement of fiber-Mo and epi-Mo sputtered at 300°C, 500°C, and 700°C (Figure S5), and the approximate thickness of the AlN and Mo layers determined using ToF-SIMS (Table S1). ToF-SIMS measurements of oxygen levels in Mo sputtered at 300°C, 500°C, and 700°C on fiber textured and epitaxial AlN (Figure S6).

\section*{Acknowledgements}
This project was performed within the framework of COMET—Competence Centers for Excellent Technologies and ASSIC Austrian Smart Systems Integration Research Center, which is funded by BMVIT, BMDW, and the Austrian provinces of Carinthia and Styria. The COMET programme is run by FFG. The authors would also like to thank Nadine Brückner for XRD measurements. Dr.Jürgen weippert, Dr.Maximilian Kessel, and Dr. Bernd Heinz for their help with the sputter tool, and fruitful discussions. 

\section*{Conflict of Interest}
The authors have no conflicts to disclose.

\section*{Data availability}
The data that support the findings of this study are available from the corresponding author upon reasonable request.

\bibliography{library.bib}

\bibliographystyle{abbrv}

\end{document}


\vspace*{0.35in}

\begin{flushleft}
{\Large
\textbf\newline{Supplementary material for Structural and electrical properties of fiber textured and epitaxial molybdenum thin films prepared by magnetron sputter epitaxy}
}
\newline
\\
Author Balasubramanian Sundarapandian\textsuperscript{1,*},
Author Mohit Raghuwanshi\textsuperscript{1},
Author Patrik Straňák\textsuperscript{1},
Author Yuan Yu\textsuperscript{1},
Author Haiyan Lyu\textsuperscript{2},
Author Mario Prescher\textsuperscript{2},
Author Lutz Kirste\textsuperscript{1}
Author Oliver Ambacher\textsuperscript{3}
\\
\bigskip
\bf{1} Fraunhofer Institute for Applied Solid State Physics, Tullastraße 72, 79108 Freiburg im Breisgau, Germany
\\
\bf{2} Institute of Physics (IA), RWTH Aachen University, Sommerfeldstraße 14, 52074 Aachen, Germany
\\
\bf{3} Institute for Sustainable Systems Engineering (INATECH), University of Freiburg, Emmy-Noether-Straße 2, University of Freiburg, 79110 Freiburg im Breisgau, Germany
\\
\bigskip
* balasubramanian.sundarapandian@iaf.fraunhofer.de

\end{flushleft}

\begin{figure}[!h]
	\centering
	\includegraphics[width=0.9\linewidth]{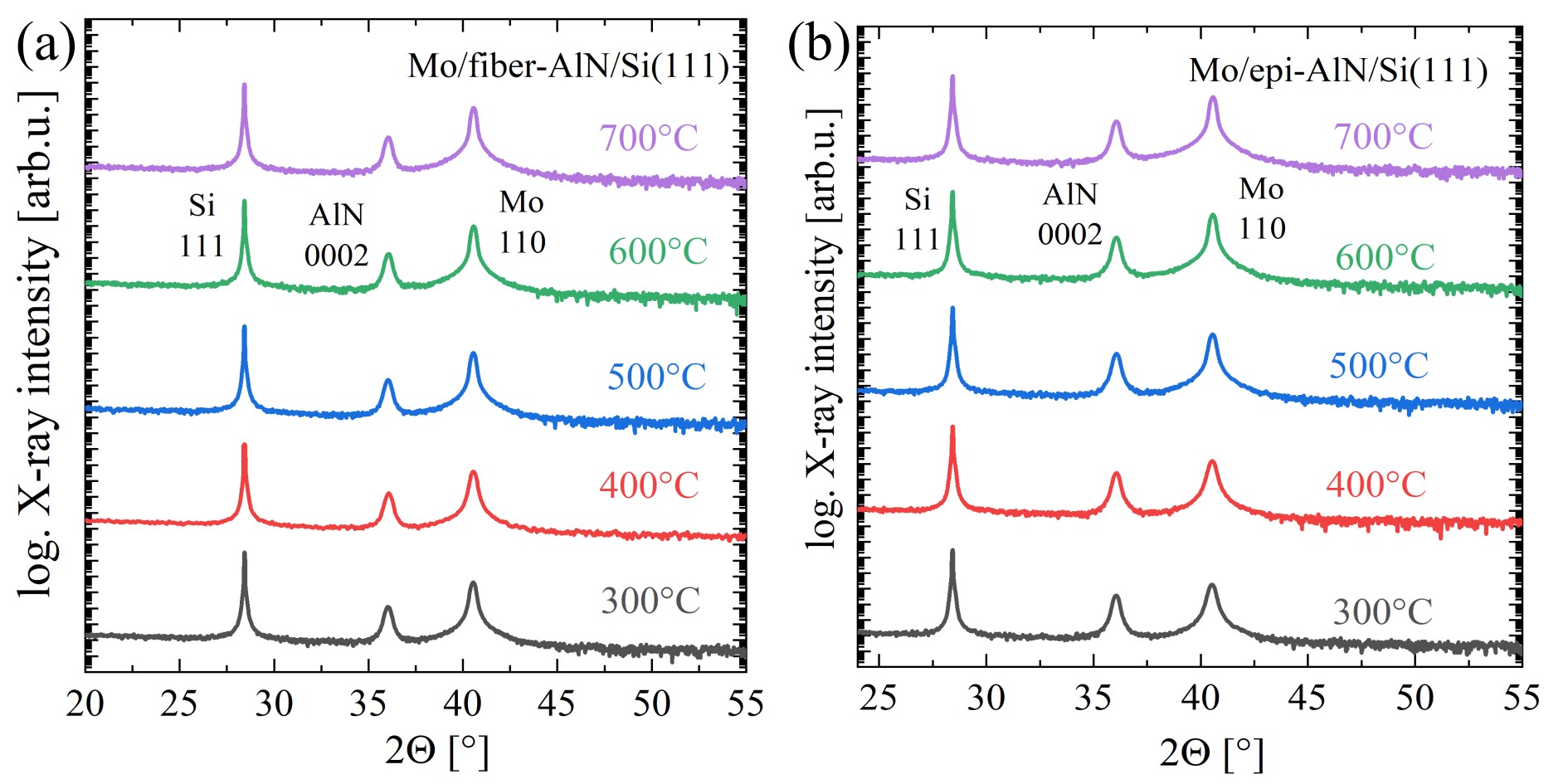}
\caption{XRD-$2\theta / \theta$ analysis of Mo grown at various temperatures on (a) fiber-textured AlN, and (b) epitaxial AlN.}
\label{sup_fig:th_2th}
\end{figure}

XRD-$2\theta / \theta$ scans of molybdenum deposited at various temperatures on fiber-textured, and epitaxial AlN are shown in figure \ref{sup_fig:th_2th} (a), and (b) respectively. The scans show absence of Mo$_3$Si peaks, showing that the formation of Mo$_3$Si is arrested by AlN seed layer, in the investigated temperature range.

\begin{figure}[!h]
	\centering
	\includegraphics[width=0.9\linewidth]{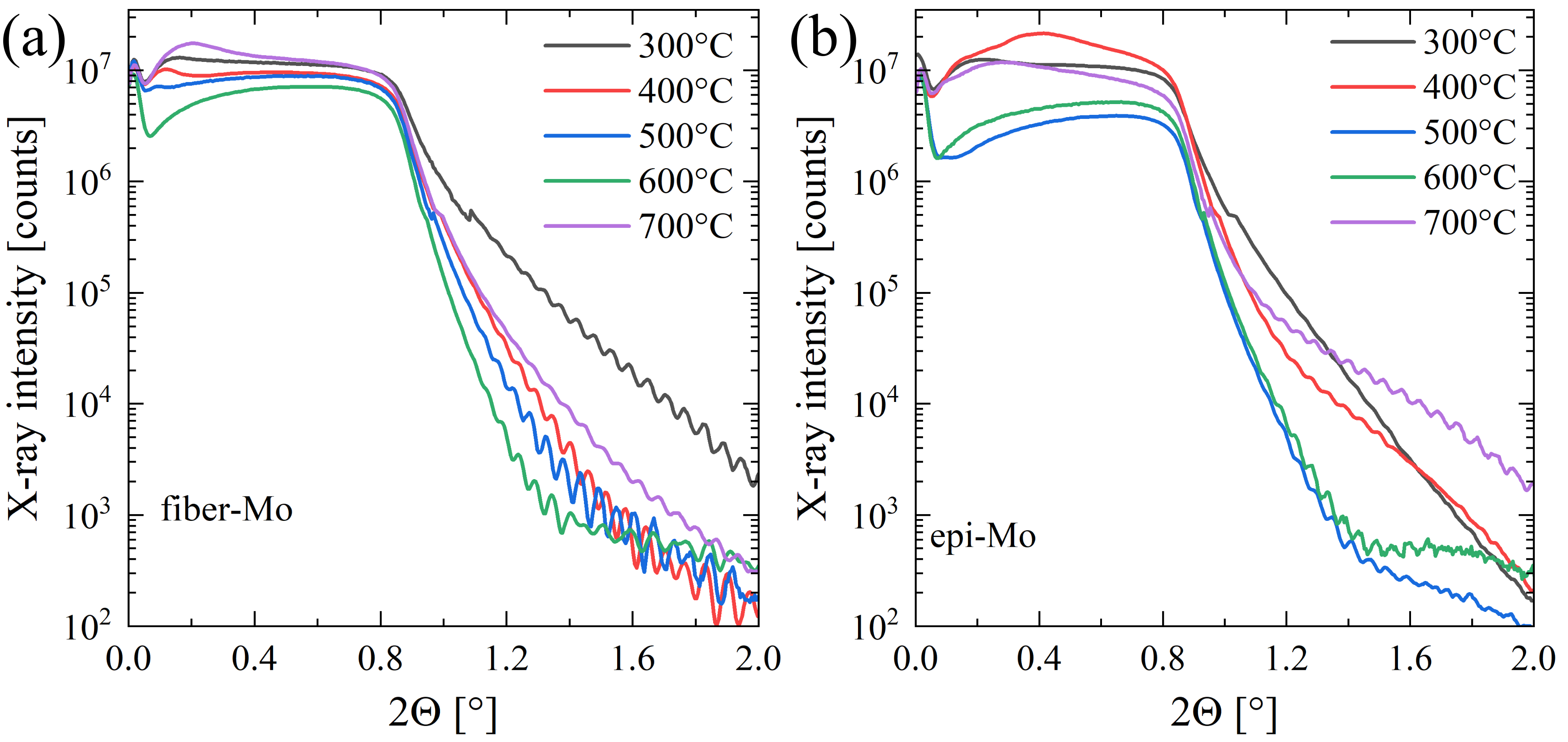}
\caption{X-Ray reflectivity profiles of (a) fiber-textured, and (b) Epitaxial Mo deposited at various growth temperatures.}
\label{sup_fig:xrr_mo}
\end{figure}

XRR recorded for fiber-textured and epitaxial-Mo at various growth temperatures are shown in figure \ref{sup_fig:xrr_mo} (a) and (b) respectively. For both fiber-textured and epitaxial Mo, the critical angle $\theta _c$ for total external reflection approximately remained constant and hence the density of Mo extracted by simulating the reflections. The rate of decay of the reflections depend on the $R_\mathrm{q}$ of the films, and in both the cases the rate of decay increases up to 600°C, and decreases at 700°C, following a similar trend as $R_\mathrm{q}$ . 

\newpage

\begin{figure}[!h]
	\centering
	\includegraphics[width=0.8\linewidth]{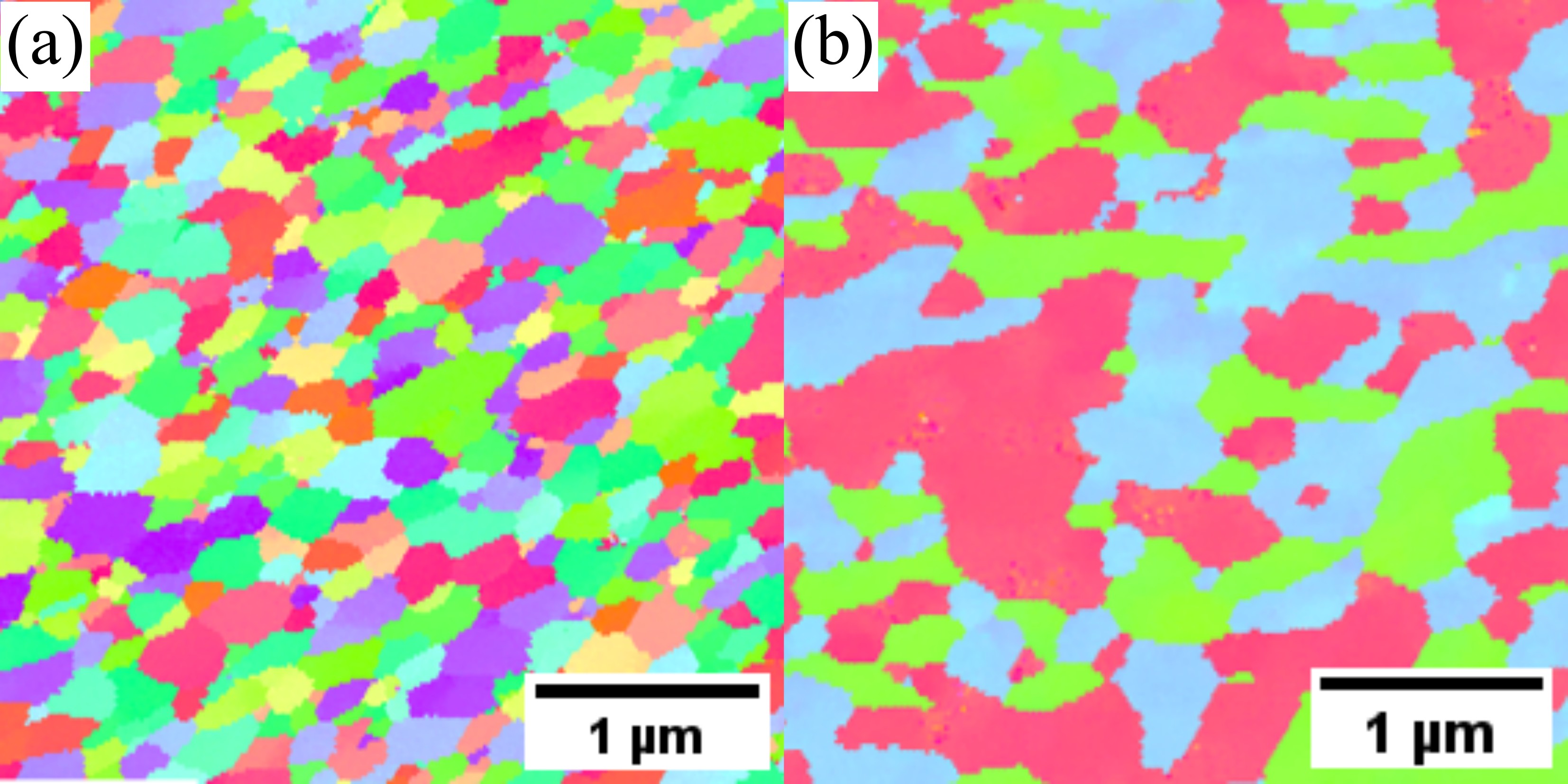}
\caption{In-plane orientation maps obtained using EBSD for Mo grown at 700°C on (a) fiber-textured AlN, and (b) epitaxial AlN}
\label{sup_fig:ebsd}
\end{figure}

Mo films sputtered at 600°C, and 700°C on fiber-textured and epitaxial AlN were analysed using EBSD to determine the grain size distribution, and orientation of grains. In-plane orientation map of Mo grown at 700°C on fiber-textured and epitxial AlN is shown in figure \ref{sup_fig:ebsd} (a) and (b) respectively. The maps clearly demonstrate that the Mo grains on fiber textured AlN is randomly oriented, while on epitaxial AlN it has three rotational domains, reinforcing the results observed through XRD. It must be noted that since all the grains of Mo are oriented out of plane along 110. A typical inverse pole figure would show no differences in the color of the grain. Hence EBSD maps shown here uses an arbitrary plane of reference to assign same color for grains with same orientation .

\begin{figure}[!h]
	\centering
	\includegraphics[width=1\linewidth]{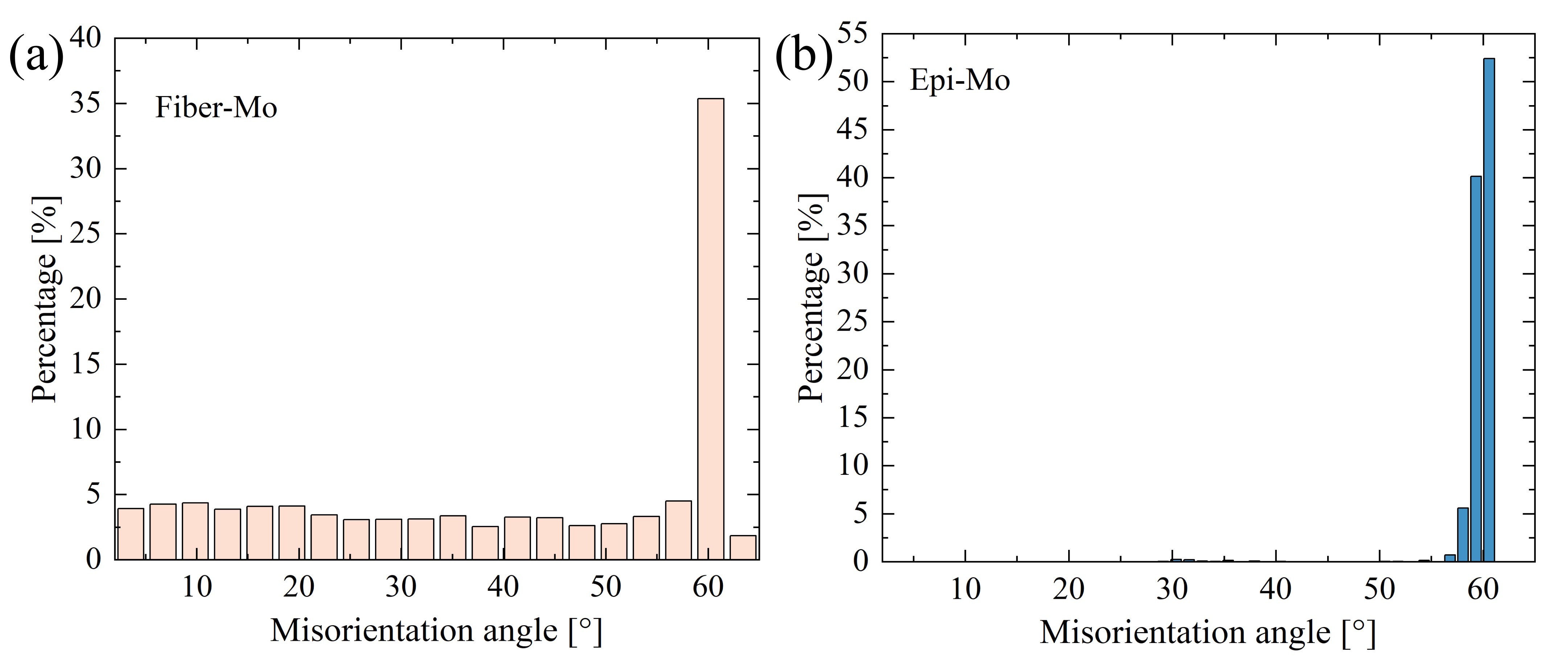}
\caption{Distribution of grain misorientation in Mo grown at 600°C on (a) fiber-textured, and (b) epitaxial AlN }
\label{sup_fig:misorientaion}
\end{figure}

The distribution of grain misorientation determined using EBSD for Mo grown at 600°C on fiber textured, and epitaxial AlN is shown in figure \ref{sup_fig:misorientaion} (a), and (b) respectively. The distributions clearly show that the Mo grains prefer to grow with a misorientation angle of 60° irrespective of their crystallographic texture. However, in fiber-textured Mo, grains with mis-orientation angle other than 60° are also present, which is absent in epitaxial Mo. \\

\begin{figure}[!h]
	\centering
	\includegraphics[width=1\linewidth]{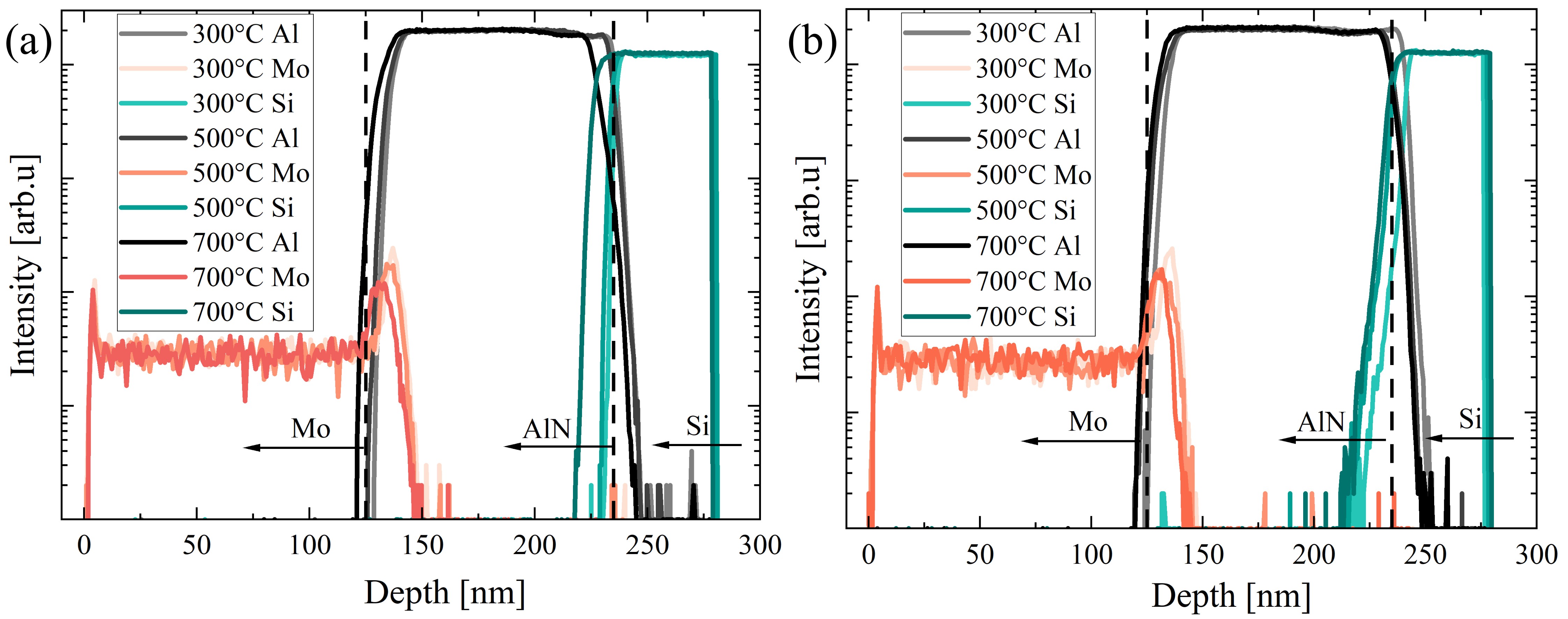}
\caption{ToF-SIMS measurements of Mo sputtered at 300°C, 500°C, and 700°C on (a) fiber textured and (b) epitaxial AlN respectively }
\label{sup_fig:sims_depth}
\end{figure}

In order to determine the thickness, growth rate, and to check whether there is diffusion of Mo and Si into AlN ToF-SIMS measurement were carried out. For this purpose, Mo samples grown at 300°C, 500°C, and 700°C on fiber textured and epitaxial AlN were chosen, and the results are shown in \ref{sup_fig:sims_depth}. The approximate thickness of AlN, and Mo layers determined using these measurements are listed in table \ref{sub_tab:thickness}. The table clearly demonstrates that the thickness of Mo on all the samples are approximately the same, indicating that the growth rate of Mo is not dependent on the growth temperature, and crystallographic texture of the AlN films. Additionally figure \ref{sup_fig:sims_depth} also demonstrates that there is slight diffusion of Mo and Si into AlN. However, the AlN layer is thick enough to prevent formation of molydenum silicides.

\begin{table}[!h]
\centering
\caption{Approximate thickness of AlN, and Mo layers determined using ToF-SIMS measurement for Mo sputtered at 300°C, 500°C, and 700°C on fiber-textured and epitaxial AlN.}
\begin{tabular}{ccc}
\hline\hline
\textbf{Sample}             & \textbf{\begin{tabular}[c]{@{}c@{}}AlN thickness\\ {[}nm{]}\end{tabular}} & \textbf{\begin{tabular}[c]{@{}c@{}}Mo thickness\\ {[}nm{]}\end{tabular}} \\ \hline
300°C Mo/fiber-textured AlN & 88                                                                        & 117                                                                      \\
500°C Mo/fiber-textured AlN & 89                                                                        & 121                                                                      \\
700°C Mo/fiber-textured AlN & 85                                                                        & 116                                                                      \\
300°C Mo/epitaxial AlN      & 96                                                                        & 120                                                                      \\
500°C Mo/epitaxial AlN      & 90                                                                        & 118                                                                      \\
700°C Mo/epitaxial AlN      & 97                                                                        & 116                                                                      \\ \hline\hline
\end{tabular}
\label{sub_tab:thickness}
\end{table}

\begin{figure}[!h]
	\centering
	\includegraphics[width=0.5\linewidth]{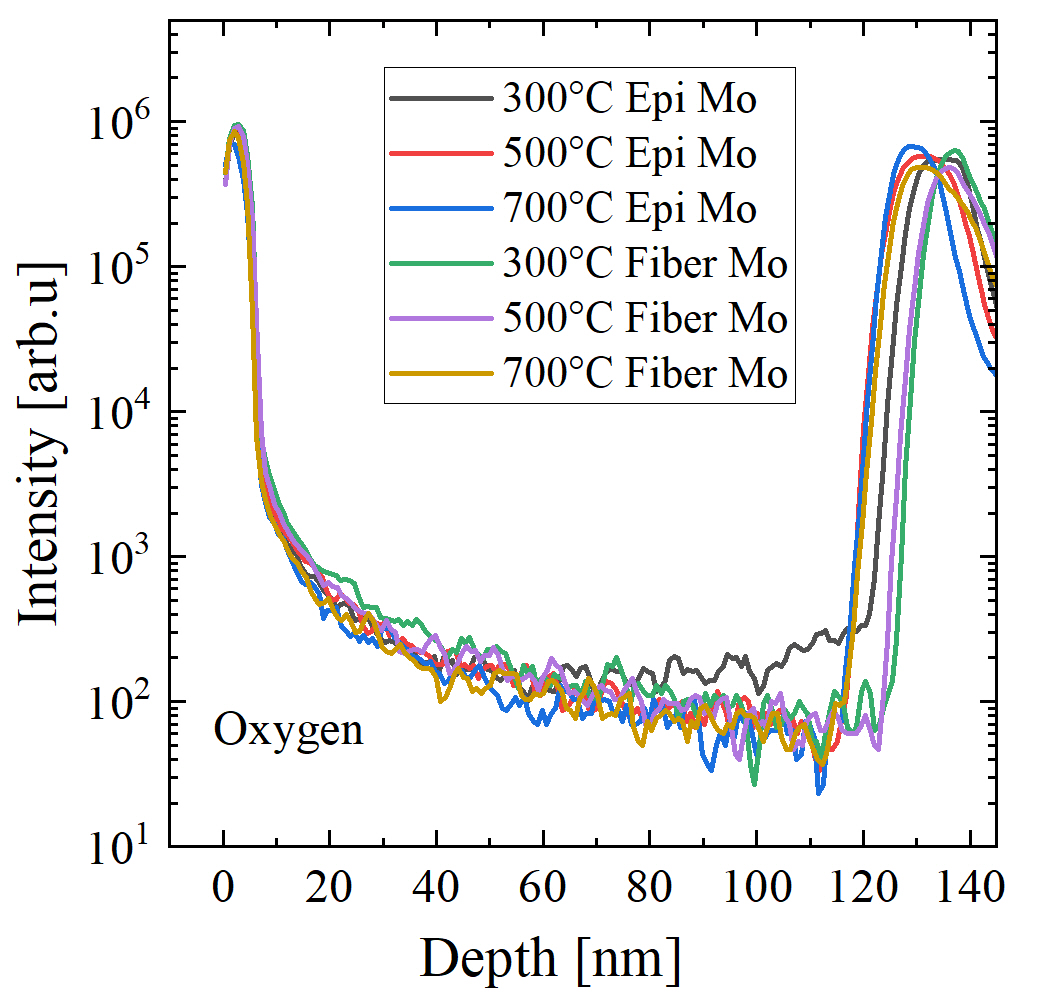}
\caption{ToF-SIMS measurements of oxygen levels in Mo sputtered at 300°C, 500°C, and 700°C on fiber textured and epitaxial AlN.}
\label{sup_fig:oxygen_level}
\end{figure}

Oxygen levels in Mo sputtered at 300°C, 500°C, and 700°C on fiber textured and epitaxial AlN were determined uing ToF-SIMS and the results are shown in figure \ref{sup_fig:oxygen_level}. The plot clearly demonstrates that growth temperature, and crystallographic texture has negligible influence on oxygen inclusion in Mo films. Therefore, oxygen inclusion in Mo films have negligible influence on the observed change in electrical resistivity of the films.

